\documentclass[aip,amsmath,amssymb,preprint]{revtex4-1}

\usepackage{graphicx}
\usepackage{amsmath,amssymb}
\usepackage{textcomp}
\usepackage{hyperref}
\usepackage{xcolor}
\usepackage{bm}

\newcommand{\dd}{\ensuremath{\mathrm{d}}}

\newcommand{\pdiff}[2]{\ensuremath{\frac{\partial {#1}}{\partial {#2}}}}
\newcommand{\vns}{\ensuremath{v_\mathrm{ns}}}
\newcommand{\vn}{\ensuremath{v_\mathrm{n}}}
\newcommand{\vs}{\ensuremath{v_\mathrm{s}}}

\newcommand{\bv}[1]{\ensuremath{\mathbf{#1}}}

\def\3He{$^3$He}
\def\4He{$^4$He}

\begin{document}

\title{Dynamics of quantum turbulence in axially rotating thermal counterflow}

\author{R. Dwivedi}
\author{T. Dunca}
\author{F. Novotný}
\author{M. Talíř}
\author{L. Skrbek}
\affiliation{Faculty of Mathematics and Physics, Charles University, Ke Karlovu 3, Prague, 121 16, Czech Republic}
\author{P. Urban}
\author{M. Zobač}
\author{I. Vlček}
\affiliation{The Czech Academy of Sciences, Institute of Scientific Instruments, CZ-61264 Brno, Czech Republic}
\author{E. Varga}
\email{emil.varga@matfyz.cuni.cz}
\affiliation{Faculty of Mathematics and Physics, Charles University, Ke Karlovu 3, Prague, 121 16, Czech Republic}

\date{\today}

\begin{abstract}
Generation, statistically steady state, and temporal decay of axially rotating thermal counterflow of superfluid $^4$He (He~II) in a square channel is probed using the second sound attenuation technique, measuring the density of quantized vortex lines. The array of rectilinear quantized vortices created by rotation strongly affects the development of quantum turbulence. At relatively slow angular velocities, the type of instability responsible for the destruction of the laminar counterflow qualitatively changes: the growth of seed vortex loops pinned on the channel wall becomes gradually replaced by the growth due to Donnelly-Glaberson instability, which leads to rapid growth of helical Kelvin waves on vortices parallel with applied counterflow. The initial transient growth of vortex line density that follows the sudden start of the counterflow appears self-similar, linear in dimensionless time, $\Omega t$. We show numerically that Kelvin waves of sufficiently strong amplitude reorient the vortices into more flattened shapes, which grow similarly to a free vortex ring. The observed steady state vortex line density at sufficiently high counterflow velocity and its early temporal decay after the counterflow is switched off is not appreciably affected by rotation. It is striking, however, that although the steady state of rotating counterflow is very different from rotating classical grid-generated turbulence, the late temporal decay of both displays similar features:  the decay exponent decreases with the rotation rate $\Omega$ from -3/2 towards approximately -0.7, typical for two-dimensional turbulence, consistent with the transition to bidirectional cascade.
\end{abstract}

\maketitle
\section{Introduction}

Rotating turbulent flows represent an important physical phenomenon occurring in a wide range of systems from industrial rotating machinery to geophysical flows\cite{Vallis2017}. Rotating flows, rather than being driven primarily by the interaction with the boundaries, are strongly affected by the Coriolis body force $\bv F_c = -2\rho\boldsymbol{\Omega}\times \bv u$, where $\boldsymbol{\Omega}$ is the angular velocity and $\bv u$ the velocity field and $\rho$ the density. In rapidly rotating flows, characterized by low Rossby number $\mathrm{Ro} = u/l\Omega$ (with $u$, $l$ being characteristic velocity and length scale, respectively), the competition between excitation of inertial waves propagating parallel to the axis of rotation and the two-dimensionalization of the flow due to the Taylor-Proudman theorem\cite{Davidson2015} leads to a phase transition-like appearance of the inverse cascade\cite{Kan2020,Kan2022a} which joins a growing body of intensely-studied phase-transition-like phenomena in fluid turbulence\cite{Cortet2010,Hof2023}. The inverse cascade can lead to the spontaneous formation of large-scale structures\cite{Boffetta2012} or an intermediate regime with an energy flux loop\cite{ClarkDiLeoni2020} with the inverse cascade arrested at an intermediate scale. In this work, we study non-classical vortex instability in a rotating quantum liquid, superfluid \4He (He~II), in order to shed light on the universal features of rotating turbulence.

The nature of the inertial waves changes drastically in He~II due to the presence of quantized vortices\cite{Tilley_book}, line-like topological defects with diameter $a\approx 0.15$~nm around which the flow circulation is restricted to a single quantum of circulation $\kappa\approx 9.98\times 10^{-8}$ m$^2$/s. In the steady-state under rotation, the superfluid becomes threaded by a hexagonal lattice of rectilinear singly quantized vortex lines\cite{Osborne1950}, and on length scales larger than the mean distance between quantized vortices, He~II mimics classical solid body rotation. The vortex line density (total length of vortices per unit volume or, equivalently, the number of rectilinear vortices per unit area) obeys the Feynman rule
\begin{equation}
    L =\frac{2\Omega}{\kappa},
    \label{eq:Feynman}
\end{equation}
where $\Omega$ is the angular velocity of rotation. Since each quantized vortex supports helical Kelvin waves \cite{Fonda2014}, the dispersion relation of waves propagating along the axis of rotation is\cite{Glaberson1974}
\begin{equation}
    \label{eq:dispersion}
    \omega = 2\Omega + \beta k^2,
\end{equation}
where $\beta = \kappa/4\pi \ln\left(b/a\right)$ and the wave-vector $\bv k$ is assumed to be oriented parallel to the axis of rotation; $b$ is the typical distance between vortices.

Above approximately 1 K, flows of superfluid helium can be described using the two-fluid model as a superposition of viscous normal flow (with temperature-dependent density $\rho_n$ and velocity $\bv v_n$) and inviscid superflow ($\rho_s$, $\bv v_s$). Historically most studied type of turbulence in He~II is the thermal counterflow\cite{Vinen1957a,Vinen1957b,Vinen1957,Vinen1958,QTbook}, which can be set up by applying heat flux $\dot q$ at a closed end of a channel with its other end open to the bath of He~II. The heat flux $\dot q$ is carried in a convective manner by the normal fluid, and by conservation of mass, a superfluid current arises in the opposite direction, and counterflow velocity is established: $\vns = \dot q/ \rho_\mathrm{s} s T$, where $T$ is the temperature and $s$ the specific entropy. For sufficiently low $\vns$, the flow of the viscous normal fluid is laminar, and there are no quantized vortices in the superfluid component (except for the remnant ones \cite{Awschalom1984}). Upon increasing $\dot q$, on exceeding the small critical counterflow velocity $\vns^\mathrm{cr}$, thermal counterflow becomes turbulent, and a tangle of quantized vortices is generated by extrinsic nucleation and reconnections. Both the superfluid component (as a tangle of quantized vortices) and the viscous normal fluid can become turbulent\cite{tough_rev,Marakov2014,Gao2017}.

The hexagonal array of vortex lines can become unstable in counterflow oriented along the axis of rotation\cite{Cheng1973,Swanson1983,Peretti2023} -- the so-called Donnelly-Glaberson instability -- which leads to the excitation of helical Kelvin waves above a critical counterflow velocity\cite{Glaberson1974,Ostermeier1975}. A similar instability is believed to be at least partly responsible for the transition to turbulence also in non-rotating thermal counterflow through the so-called vortex mill mechanism\cite{Schwarz1990}.
Experiments\cite{Swanson1983} in rotating counterflow at small velocities and vortex filament numerical simulations\cite{Tsubota2003,Tsubota2004} showed the suppression of the critical velocity for the turbulent growth of vorticity to very small values for rapidly rotating turbulence and possibly a second critical velocity connected to the development of disordered tangle. Recently, the excitation of vortex waves was directly visualized\cite{Peretti2023} where, however, the waves were not forced by Donnelly-Glaberson instability but rather by oscillatory counterflow.

In this work, we experimentally study rotating counterflow probed by the second sound attenuation. We confirm the results of Swanson \textit{et al.}\cite{Swanson1983} and extend these results to the dynamics of turbulence growth and decay. With complementary numerical simulations, we relate the initial growth of vortex line density to Kelvin wave growth on individual vortices in the vortex lattice and show that rotating quantum turbulence decays quasi-classical at long decay times consistent with rotating turbulence in elongated domains \cite{Morize2006,Kan2020}.

The paper is organized as follows: The experimental setup and protocol are described in Sec.~\ref{sec:setup}, and experimental results on steady-state turbulence are shown in Sec.~\ref{sec:steady}. Dynamical transitional behavior is addressed both experimentally and numerically in Sec.~\ref{sec:transient}, after which discussion and conclusions follow.

\section{Experimental setup}
\label{sec:setup}

\begin{figure}
    \centering
    \includegraphics{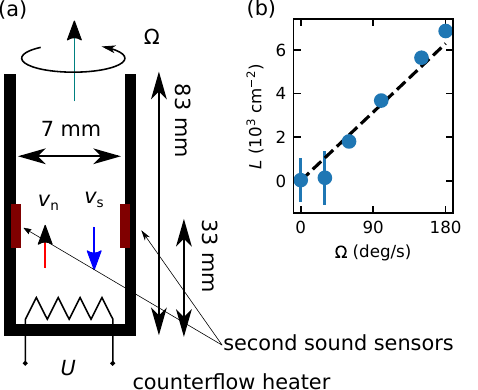}
    \caption{(a) Schematic diagram of the $7\times 7$~mm square cross-section counterflow channel. The channel axis is aligned with the axis of rotation. The resistive heater placed at the bottom generates the counterflow of normal and superfluid components. (b) Vortex line density measured at $T=1.65$~K under steady rotation using second sound attenuation without the presence of thermal counterflow. The vortex line density was calculated using \eqref{eq:L-polarized}, and the dashed line shows the Feynman rule \eqref{eq:Feynman}}
    \label{fig:setup}
\end{figure}

The geometry of the counterflow channel used in the experiment is illustrated in Fig.~\ref{fig:setup} and is similar to the past experiments of Swanson \emph{et al.}\cite{Swanson1983}. The brass flow channel with a square cross-section $A = 7 \times 7$~mm$^{2}$ and length $H = 83$~mm is placed vertically on the axis of rotation of the cryostat. A resistive wire heater ($R \approx 16$~\textohm) glued to a flat surface is inserted inside the channel from the bottom to provide the counterflow heat flux, keeping the other end open to the liquid helium bath.

The vortex line density in the channel is detected using second sound attenuation \cite{Varga2019}. The channel acts as a semi-open acoustic resonator for the second sound, which is excited and detected at approximately 24 kHz by a pair of oscillating membrane ($<1$~\textmu m pore size) second sound transducers \cite{Zimmermann1986}.

The change in the attenuation of the resonance can be used to calculate the vortex line density $L$. Here, however, a complication exists since the attenuation due to vortices depends on the angle $\theta$ between the vortex and direction of sound propagation as $\sin^2\theta$, i.e., the second sound is not dissipated when propagating parallel to a straight vortex \cite{Babuin2012}. For a disordered, turbulent tangle, one can assume a random orientation of vortices uniformly distributed in all directions, which results \cite{Babuin2012} in
\begin{equation}
    \label{eq:L-random}
    L=\frac{6\pi\Delta_0}{B\kappa}\left(\frac{a_0}{a}-1\right),
\end{equation}
where $B$ is a temperature-dependent mutual friction parameter \cite{Donnelly1998}, $\Delta_0$ is the width of the unattenuated second sound resonance, and $a_0$, $a$ are the amplitudes on resonance of unattenuated and attenuated peaks, respectively. However, in a situation when all vortices are oriented perpendicular to the direction of sound propagation (i.e., simple rotation without counterflow in the present experiment), on average, $\sin^2\theta = 1$ and the expression for vortex line density is instead
\begin{equation}
    \label{eq:L-polarized}
    L=\frac{4\pi\Delta_0}{B\kappa}\left(\frac{a_0}{a}-1\right),
\end{equation}
which differs from \eqref{eq:L-random} by a factor of 2/3. For the intermediate cases of partially polarized turbulent tangles, the appropriate expression will interpolate between these two extremes. Since the polarization of the tangle cannot be determined in the present experiment, one should take into account an additional systematic uncertainty up to 33\%  on the value of vortex line density in addition to the statistical uncertainties shown in Sec.~\ref{sec:results} below.

The typical kinetic and thermal response time\cite{Varga2018} of the present experiment are of the order of 10~ms. The second sound response time (inverse linewidth of the resonance) is, at most, approximately 40~ms, which ensures that none of the transient effects studied, which occur on the timescale of seconds, are significantly affected by the non-turbulent system relaxation times.

The experimental setup was mounted on a custom-made rotating platform, described in detail in Appendix A. In this work, the rotation of the cryostat was restricted to a maximum angular speed of 180 degrees per second.

\section{Results}
\label{sec:results}

To verify the performance of our experimental setup, we first measured the vortex line density in steady rotation, which ought to obey Feynman's rule \eqref{eq:Feynman}. The vortex line density determined from the frequency sweeps of the second sound resonance is shown in Fig.~\ref{fig:setup}(b) for zero counterflow heat flux and $T=1.65$~K. There is good agreement between theoretical expectation \eqref{eq:Feynman} and observed vortex line density, suggesting that negligible turbulence is generated by rotation alone. Note that for the case of simple rotation, all vortices are expected to be aligned parallel to the axis of rotation and perpendicular to the direction of propagation of the second sound, therefore vortex line density is calculated using \eqref{eq:L-polarized} rather than \eqref{eq:L-random} which is appropriate for the turbulent cases.

Due to a slight overheating of the flow channel from the heat flux dissipated by the counterflow heater, we restricted our study to 1.65~K, where the second sound velocity has a local maximum\cite{Donnelly1998}. This significantly reduces the shifting of the second sound resonance, which can result in a spurious decrease in signal strength that can be erroneously interpreted as an increase in vortex line density. This is especially relevant to the transient effects, where it is not possible to sample the full resonance curve. Note that this overheating occurs inside the the flow channel, the bath temperature is stabilisied to approximately 1~mK regardless of the heat dissipated in the channel.

\subsection{Steady-state thermal counterflow}
\label{sec:steady}

\begin{figure}
    \centering
    \includegraphics{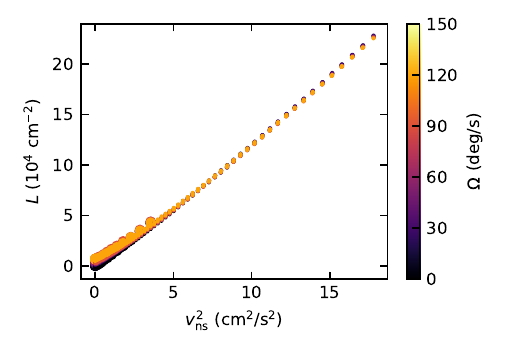}
    \caption{Vortex line density in steady state turbulence as a function of counterflow velocity for several rotation speeds of the cryostat. Larger points (at small velocities) are measured using full frequency sweeps, and smaller points (extending to higher velocities) using a measurement at fixed frequency and stepped counterflow velocity.}
    \label{fig:steady-state}
\end{figure}

\begin{figure}
    \centering
    \includegraphics{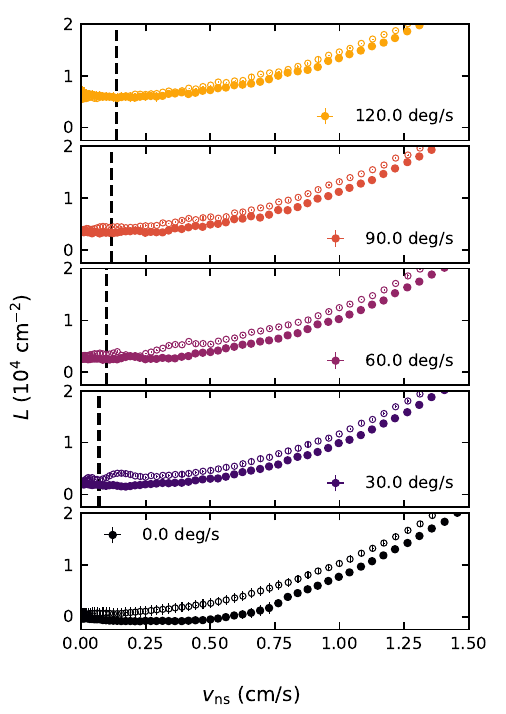}
    \caption{Vortex line density hysteresis for different rotation speeds. Vortex line density was measured at a fixed second sound frequency with step-wise increasing (filled points) or decreasing (empty points) counterflow velocity. Near the critical velocity, pronounced hysteresis is observed, which is suppressed in the presence of fast rotation. Note that the vortex line density in the $v_\mathrm{ns}\to 0$ limit is higher than the Feynman rule \eqref{eq:Feynman} by a factor of 3/2 due to the use of the random orientation expression \eqref{eq:L-random}. The vertical dashed lines are the critical velocities \eqref{eq:critical-velocity} given by Glaberson \emph{et al.}\cite{Glaberson1974}.}
    \label{fig:hysteresis}
\end{figure}

The steady-state vortex line density as a function of counterflow velocity for several rotation speeds is shown in Fig.~\ref{fig:steady-state} and Fig.~\ref{fig:hysteresis}. In this case, the expression \eqref{eq:L-random} assuming random vortex orientations for calculating vortex line density was used. For high counterflow velocities, the expected $L\propto v_\mathrm{ns}^2$ scaling is reproduced, which occurs in flows with forced nonzero $\vns$ \cite{Babuin2012} regardless of the angular velocity of the system. In agreement with Swanson \emph{et al.} \cite{Swanson1983}, the vortex-free state is suppressed as the rotation rate increases. The initial critical velocity of the rotating vortex array was estimated by Glaberson \emph{et al.}\cite{Glaberson1974} similarly to the Landau critical velocity as 
\begin{equation}
    \label{eq:critical-velocity}
    v_\mathrm{c} = 2\sqrt{2\Omega\beta},
\end{equation}
with $\beta$ defined under \eqref{eq:dispersion}, which agrees in the order of magnitude with transitions shown in Fig.~\ref{fig:hysteresis} (vertical dashed lines).

Additionally, the counterflow velocity (heat flux) in Fig.~\ref{fig:hysteresis} was increased or decreased in steps without being interrupted. At zero rotation, pronounced hysteresis exists between the increasing and decreasing direction of the heat stepping, which is gradually suppressed as the rotation rate increases. In particular, the sharp transition in the increasing heat flux data is either suppressed or absent. This is generally in agreement with the picture that the nature of the transition to turbulence changes even in slowly rotating systems -- rather than the growth starting from a random distribution of vortex loops pinned on the channel walls\cite{Varga2017a,Pomyalov2020}, the flow is linearly unstable at essentially all velocities due to the Donnelly-Glaberson instability\cite{Swanson1983,Tsubota2003,Tsubota2004}. Note that the slight offset between the increasing and decreasing velocity at sufficiently high velocities is most likely due to the relaxation of the tangle density and finite heat flux ramp rate.

\subsection{Transient behavior}
\label{sec:transient}

\begin{figure}
    \centering
    \includegraphics{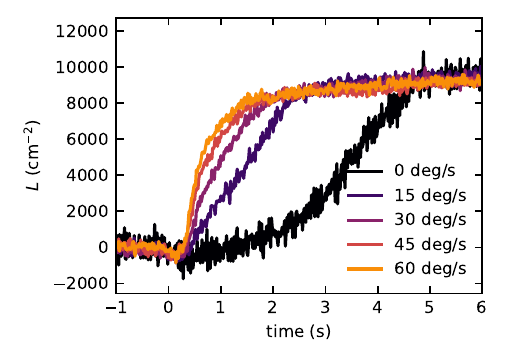}
    \caption{Initial transient growth of vortex line density. At $t=0$, the counterflow velocity is increased suddenly from $v_\mathrm{ns} = 0$ to $v_\mathrm{ns} \approx 0.9$ cm/s. For each rotation speed, the transient was measured in 20 realizations, shown are the averages. The second sound signal was measured with a fixed sampling of 64 Hz. For the reference amplitude $a_0$, the average amplitude for $t < 0$ was used. Therefore, all curves start at $L=0$.}
    \label{fig:transient}
\end{figure}

\begin{figure}
    \centering
    \includegraphics{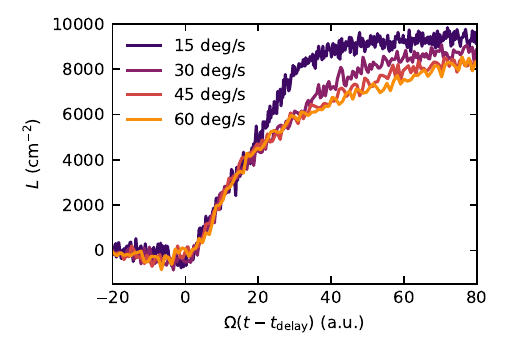}
    \caption{Normalized initial transient growth of vortex line density. As Fig.~\ref{fig:transient} with time axis normalized by the rotation speed and shifted by $t_\mathrm{delay}$ = 0.25~s.}
    \label{fig:transient-normalized}
\end{figure}

The initial growth of the vortex line density after the flow is suddenly switched on (at $t=0$) is shown in Fig.~\ref{fig:transient} for several rotation speeds. Even the lowest rotation speed studied results in a severe change of the transition behavior, again in line with the expectation that the type of instability is altered by the presence of axially oriented vortex lines. Figure~\ref{fig:transient-normalized} shows the same data with the time axis normalized by the angular velocity. All rotation speeds show a brief rotation-independent delay, after which linear in time growth starts. The linear regime is self-similar for different non-zero angular velocities as all data show the same growth rate in dimensionless time $\Omega t$.

\begin{figure}
    \centering
    \includegraphics{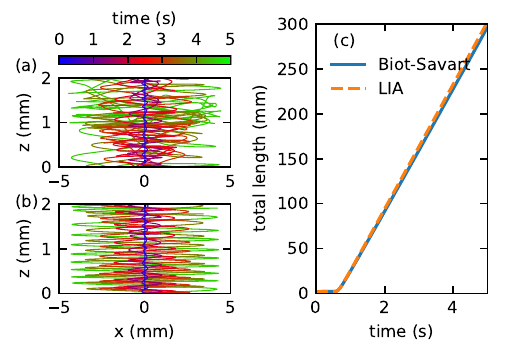}
    \caption{Vortex filament simulation of the growth of a Kelvin wave on an isolated vortex in $v_\mathrm{ns} = 1$~cm/s axial counterflow (along the $z$ axis) at 1.65 K. (a,b) Snapshots of the vortex at several times from 0 to 5 s. The motion of the vortex was calculated using the (a) full Biot-Savart integral or (b) local induction approximation. As the initial condition, a helical Kelvin wave with an amplitude of 1 nm and wavelength of 1 mm was used. (c) The total length of the vortex as a function of time for the two calculations.}
    \label{fig:KW-growth-sim}
\end{figure}

The origin of the linear growth regime is in the growth of Kelvin waves on individual vortices. A vortex filament simulation \cite{Hanninen2014} (see appendix for the details on the simulations) of an isolated vortex line oriented parallel to counterflow of $v_\mathrm{ns} = 1$~cm/s at 1.65 K is shown in Fig.~\ref{fig:KW-growth-sim}. The boundary conditions are open in the directions perpendicular to the counterflow direction ($x$ and $y$) and periodic along the counterflow direction ($z$, 1 mm box size). Once the initial helical perturbation grows to a sufficiently large amplitude, the radial expansion of the vortex is mostly due to the imposed counterflow, similar to a planar vortex ring. The approximately linear growth of the helix amplitude was also shown analytically for large times by Van Gorder \cite{VanGorder2015a}. The full non-local interaction, shown in Fig.~\ref{fig:KW-growth-sim}(a), results in a more complex vortex shape than the local induction approximation, shown in Fig.~\ref{fig:KW-growth-sim}(b), however, the growth of the total length is essentially unaffected as can be seen in Fig.~\ref{fig:KW-growth-sim}(c). The total length of each individual vortex in the array, therefore, grows as $L_1 \propto t$ and the number of vortices in the channel $n_V\propto \Omega$. It follows that the total vortex line density $L \propto \Omega t$ as seen in Fig.~\ref{fig:transient-normalized}. As a corollary, the end of the linear growth regime is unlikely to be the result of nonlinear Kelvin wave interactions on individual vortices but rather must occur once the interaction between neighboring vortices in the initial vortex array becomes significant.

\begin{figure}
    \centering
    \includegraphics{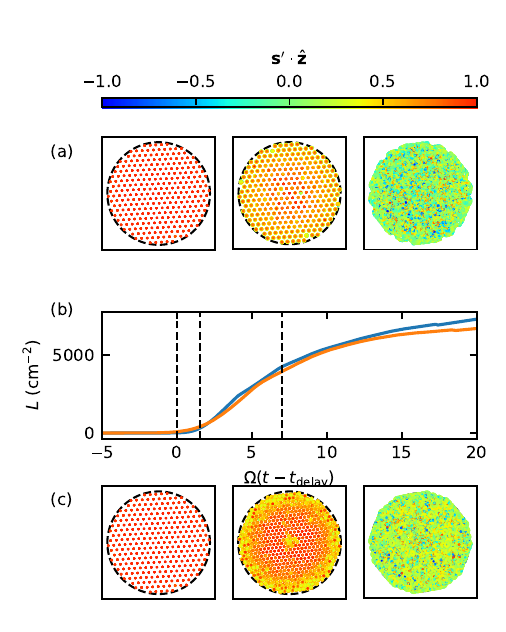}
    \caption{Simulation of axial counterflow of $\vns = 1$~cm/s applied to a rotating lattice of vortices at 1.65 K. (a,c) Snapshots of the vortex tangle for 15 deg/s, (a), and 30 deg/s, (c), rotation at times indicated in (b) by vertical dashed lines. The color indicates the directional cosine of the local vortex tangent $\bv s'$ with the $z$-axis. The black dashed circle indicates the 4~mm radius of the vortex removal cutoff. (b) Vortex line density for 15 deg/s and 30 deg/s rotation normalized as in Fig.~\ref{fig:transient-normalized}.}
    \label{fig:rotating-cf-sim}
\end{figure}

To verify this picture, we simulated an axial counterflow with a hexagonal array of quantized vortices as the initial conditions corresponding to 15 deg/s (vortex lattice constant 0.44~mm; supplementary video 1) and 30 deg/s (0.31~mm; supplementary video 2). The vortex motion was calculated in the laboratory frame of reference, i.e., the imposed normal fluid velocity contains the corresponding solid-body rotation component. Proper treatment of solid walls in a rotating system is rather challenging (due to, i.e., corner vortices) and outside the scope of the present work, therefore we simplified the simulation to open boundary conditions with the removal of vortices, which are fully outside of a cylinder with radius $r_\mathrm{cutoff} = 4$~mm and centered on the axis of rotation. Along the $z$-axis, periodic boundary conditions with 1~mm box size were used. The snapshots of the developing tangle and the growth of the vortex line density normalized similarly to Fig.~\ref{fig:transient-normalized} are shown in Fig.~\ref{fig:rotating-cf-sim}. Similarly to the experimental case, we observe an approximately self-similar linear increase in time of the vortex line density, although the rate of growth (i.e., the tangent of $L(\tilde t)$, $\tilde t = \Omega t$) is about twice as large for the numerical simulation than for the experiment. This is likely a numerical artifact since, as can be seen in Fig.~\ref{fig:rotating-cf-sim} (a,c), the growth of the helical Kelvin waves is not uniform throughout the lattice but is more rapid toward the edges. This is most likely due to the artificial treatment of the boundary. Since the number of vortices in the simulations grows rather rapidly, we were unable to reach a steady state due to high computational cost. However, the steady-state that would likely be reached in the simulation would be rather different from the experimental case due to the artificial bounding of the tangle in the in-plane directions and is thus not particularly relevant for comparison with the experiment.

\begin{figure}
    \centering
    \includegraphics{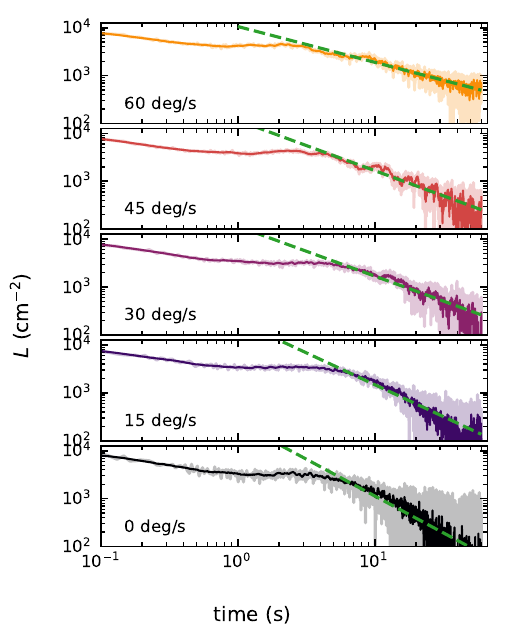}
    \caption{Decaying vortex line density. At $t=0$, the counterflow velocity is suddenly decreased from $v_\mathrm{ns} = 0.89$ cm/s to $v_\mathrm{ns} = 0$ measured in the same acquisition as the initial transients in Fig.~\ref{fig:transient}. The shown data are averaged over multiple realizations and filtered (Savitzky-Golay filtering, window size 21, and polynomial order 3). The light-colored curves in the background show unfiltered data. The green dashed lines show fits to \eqref{eq:L-decay} for $t>5$~s. The resulting decay exponents are shown in Fig.~\ref{fig:decay-exponents}}
    \label{fig:decay}
\end{figure}

\begin{figure}
    \centering
    \includegraphics{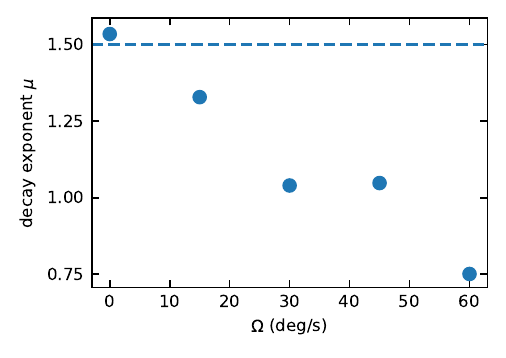}
    \caption{Vortex line density decay exponents obtained by fitting \eqref{eq:L-decay} to the decay data in Fig.~\ref{fig:decay} for $t > 5$~s. The dashed line shows the quasi-classical $\mu=1.5$ \cite{Stalp1999}.}
    \label{fig:decay-exponents}
\end{figure}

The decay of vortex line density after a sudden decrease of counterflow velocity to zero (at $t=0$) is shown in Fig.~\ref{fig:decay} for several rotation rates. The same normalization for the calculation of the vortex line density from second sound attenuation is used as for the initial transient, therefore all curves decay towards $L=0$ rather than the stationary $L$ given by \eqref{eq:Feynman}. The initial decay up to 1 - 2 s appears unaffected by rotation, and late-time decay follows a power law of the form
\begin{equation}
    \label{eq:L-decay}
    L = at^{-\mu};
\end{equation}
the decay exponent $\mu$, shown in Fig.~\ref{fig:decay-exponents}, is altered for rapid rotation from the $\mu\approx3/2$ characteristic of the quasi-classical decay \cite{Stalp1999,Skrbek2000} to approximately $\mu\approx0.75$ (although we note that the power-law behavior for the non-rotating data is only approximate and the decay exponent depends on the fitting range). For rotating decaying classical turbulence Morize and Moisy \cite{Morize2006} proposed an interpolating turbulent energy spectrum $E(k) = C_p\Omega^{(3p-5)/2}\varepsilon(t)^{(3-p)/2}k^{-p}$ ($p=5/3$ for zero rotation) based on dimensional analysis. Using the quasi-classical approximation of superfluid vorticity $|\omega| = \kappa L$, the exponent of the energy spectrum can be related \cite{Stalp1999,Skrbek2000} to the vorticity decay exponent as $\mu = 1/(p - 1)$. The decay exponent $\mu=0.75$ thus indicates an energy spectrum exponent $p\approx 2.3$, which is in good agreement with the rapidly rotating decaying classical turbulence \cite{Morize2005a}.

We also note that the number of coherent vortices in classical two-dimensional turbulence was observed to decay with exponent $\mu\approx0.7$ \cite{Tabeling2002}. Two-dimensionalization of rapidly rotating turbulence is a well-established property of classical turbulence \cite{Davidson2015}. Two-dimensional oscillatory quantum turbulence in confined geometries is known to display rather complex transitional behavior with multiple stable states of the large-scale flow \cite{Varga2020,Novotny2024b}, which are not present in the current data suggesting a simpler structure of the flow. The Rossby number defined using the quasi-classical energy dissipation rate\cite{Stalp1999} $\varepsilon = \nu_\mathrm{eff}(\kappa L)^2$ with the effective viscosity\cite{Babuin2013,Babuin2014} $\nu_\mathrm{eff}\approx0.2\kappa$ is $\mathrm{Ro} = (\varepsilon\ell)^{1/3}/(D\Omega) \approx 0.08$ for $\Omega = 60$~deg/s in Fig.~\ref{fig:decay} ($\ell = L^{-1/2}$) suggesting that the flow will be strongly affected by Coriolis forces. Assuming that the energy injection scale is close to the intervortex distance $\ell \approx 10^{-2}$~cm, for the 60 deg/s decay the dimensionless parameter in the steady state $\lambda =\ell/H \mathrm{Ro} \approx 0.033 > \lambda_c \approx 0.03$ which was shown by van Kan and Alexakis \cite{Kan2020} to be a critical value for transition to bidirectional cascade, suggesting that for $\Omega > 60$~deg/s inverse cascade is present already in the steady state.

\section{Discussion and Conclusions}
We have shown that the array of vortex lines created by rotation strongly affects the development of quantum turbulence in axially oriented thermal counterflow. The type of instability primarily responsible for the destruction of the laminar flow qualitatively changes at even relatively slow angular velocities from, presumably, the growth of random vortex loops pinned on the inner surfaces of the counterflow channel to the Donnelly-Glaberson instability which leads to rapid growth of helical Kelvin waves on vortices parallel with applied counterflow. This type of transition to turbulence no longer relies on the stochastic distribution of pinned seed vortices, which is likely the reason for the suppression hysteresis at small counterflow velocities seen in Fig.~\ref{fig:hysteresis}.

Further, we find that the initial transient growth of vortex line density after the sudden start of the counterflow proceeds in self-similar linear-in-time $L\propto \Omega t$, which differs quite strongly from the exponential growth expected for linear instability. Numerically, we have shown that Kelvin waves of sufficiently strong amplitude (mostly local) nonlinear interactions reorient the vortex into a more flattened shape which grows similarly to a free vortex ring. This free vortex ring regime is observed numerically also for an array of rotating vortices, although with a total rate of growth of vortex line density about twice the experimentally observed rate. This suggests either improperly accounted-for wall effects or the presence of normal fluid turbulence.

Finally, the observed steady-state vortex line density at sufficiently high counterflow velocities is not strongly affected by the rotation, nor is the early stage of decay after the counterflow is suddenly switched off. However, rather than quasi-classical decay characterized by $L\propto t^{-3/2}$ typically observed for the decay of strongly excited counterflow turbulence, the exponent decreases towards approximately -0.75 which is in good agreement with rotating classical decaying turbulence \cite{Morize2005a,Morize2006} and is possibly related to the decay of the number of coherent vortices in classical two-dimensional turbulence\cite{Tabeling2002}.

Despite the fact that thermal counterflow is a type of turbulence specific to superfluid helium, the naturally small forcing length scale set by the intervortex distance enables the study of rotating turbulence in new parameter regimes, helping in the understanding the universal features of rotating turbulence. While further experiments are required (especially higher rotation rates and counterflow velocities, which were not possible in the current experimental run), rotating counterflow turbulence might prove a valuable system capable of achieving flow characteristics (e.g., the ratio of forcing scale to system size) comparable to large-scale geophysical flows.

\begin{acknowledgments}
The work was supported by Charles University under PRIMUS/23/SCI/017. We are grateful to D. Schmoranzer for assistance with the initial construction of the rotating platform; B. Vejr for the machining of the experimental cell, and J. Boh\'{a}\v{c} and D. Nazarenko for ensuring liquid helium supply. 
\end{acknowledgments}

\section*{Declarations}
The authors have no conflict of interest to declare.


\appendix
\section{Rotating platform}

We use a custom-made rotating platform that supports an experimental cryostat as well as all necessary electronic equipment (see ref.\cite{Novotny2024} for a photograph). The platform rests on a central bearing with mains power for the experimental setup provided through a slip ring and the bath pump connected via a KF50 rotating feedthrough attached to the platform support structure above the cryostat. The vacuum feedthrough is double-sealed with an additional sealed compartment pressurized with \4He gas slightly above atmospheric pressure, separating room air from the low-pressure helium bath to avoid contamination due to possible leaks in the rotating rubber seals.

The rotation is driven by a 1.5 kW permanent magnet synchronous brushless three‐phase AC stepper motor powered by a variable frequency drive controlled by a programmable logic controller. The motor speed is reduced by a gearbox and flat belt transmission, which also serves as an emergency torque clutch. The available output torque ensures a braking capability from the maximum velocity of 60 RPM to a full stop in less than a second.

\section{Vortex filament simulations}

The simulations were performed using a one-way coupled vortex filament model\cite{Hanninen2014}, in which the movement of the vortices is calculated as the response to the prescribed arbitrary normal fluid velocity and potential superflow.

The vortices are represented as spatial curves discretized to a series of linked points with spacing dynamically maintained in the range of 0.025 to 0.05 mm. The motion of the individual vortex points is calculated using 4th-order Runge-Kutta time stepping with a fixed time step of 1 ms. The equation of motion of the points on the vortices ${\bv s}$,
\begin{equation}
    \label{eq:vortex-eom}
    \pdiff{\bv s}{t} = \vs + \alpha\bv s'\times(\vn - \vs) - \alpha'\bv s'\times(\bv s' \times (\vn - \vs),
\end{equation}
where $\vn$ is the prescribed normal fluid velocity and
\begin{equation}
    \label{eq:biot-savart}
    \vs(\bv s) = \beta\bv s'\times\bv s'' + \frac{\kappa}{4\pi}\int_{\mathcal L'}\frac{\dd\bv x \times (\bv s - \bv x)}{|\bv s - \bv x|^3},
\end{equation}
where $\beta = \kappa/4\pi\log(2\sqrt{l_+l_-}/a_0\sqrt{e})$ with $l_\pm$ the distances to neighboring discretization points on the vortex and the integration runs over the full tangle of vortices with the exception of the immediate vicinity of $\bv s$. For the simulations of Kelvin wave growth on a single vortex, Fig.~\ref{fig:KW-growth-sim}, the integral in \eqref{eq:biot-savart} was calculated directly (for the local induction approximation, LIA, the integral is neglected). For the rotating counterflow simulations, Fig.~\ref{fig:rotating-cf-sim}, the integral was approximated using a Barnes-Hut (tree) algorithm with the expansion of the velocity induced by the tree nodes up to quadrupolar terms.

Sections of vortices that approach each other closer than 0.25 mm are reconnected if the reconnection leads to a decrease in total length. For all simulations, $\alpha = 0.111$ and $\alpha' = 1.437\times 10^{-2}$ were used, corresponding to 1.65 K \cite{Donnelly1998}. The full code of the simulation is open-source and available at \url{https://bitbucket.org/emil_varga/openvort}.

%

\end{document}